\documentclass[traditabstract]{aa}

\makeatletter

\usepackage{pdflscape}
\makeatother

\usepackage[usenames, dvipsnames]{color} 
\usepackage{color}
    \usepackage{textcomp}
\begin{document}

\title{Correlation between UV resilience and wavelength of narrow diffuse interstellar bands} 
\titlerunning{Correlation between UV resilience and wavelength of narrow DIBs}

\author{A. Omont\inst{1}
\and H.\ F.\ Bettinger\inst{2}
}
\institute{Sorbonne Universit\'e, UPMC Universit\'e Paris 6 and CNRS, UMR 7095, Institut d'Astrophysique de Paris, France\\
\email{omont@iap.fr}
\and  Institut für Organische Chemie, Universität Tübingen, Auf der Morgenstelle 18, 72076 Tübingen, Germany\\
\email{holger.bettinger@uni-tuebingen.de}
}

\abstract{
Carriers of diffuse interstellar bands (DIBs) still need to be identified. 
In a recent paper, we reported a correlation between the DIB wavelength and the apparent UV resilience (or boost) of their carriers. 
We proposed that this might be an indication of the  important role of conjugated elongated molecules among the DIB carriers.
The aim of this paper is to further understand the origin of this correlation. The analysis of  509 
optical  DIBs on the lines of sight of HD\,183143 and/or HD\,204827 reported in the literature  shows that this correlation mainly implies the 386 narrow DIBs with a band width  $<$\,1.1\,\AA , which include most of the identified DIBs of the C$_2$ and $\zeta$ families, while the majority of  the 123  broader DIBs, including the  identified $\sigma$ DIBs, do not display such a correlation. We present a possible origin of this correlation from very strong bands of  large conjugated elongated  molecules, such as carbon chains, polyacenes, or other catacondensed polycyclic aromatic hydrocarbons.  The total amount of carbon contained in all the carriers of these narrow DIBs is a very small fraction of the interstellar carbon if their oscillator strengths are $\ge$\,1. The amount of carbon  locked in the carriers of the broader DIBs is higher, especially if their oscillator strengths are significantly weaker.
}


\maketitle 

\section{Introduction}
Identification of carriers of the diffuse interstellar bands (DIBs) remains an outstanding problem in astrophysics (see e.g.\ 
Herbig 1975, 1995; Sarre 2006; Tielens \& Snow 1995; and especially Cami \& Cox 2014; see also e.g.\ Omont, 2016; Fan et al.\ 2017, 2019; Cami et al.\ 2018 for more recent references), and to this day, only five bands have been attributed to an identified carrier, C$_{60}^+$ (Foing \& Ehrenfreund 1994; Campbell et al.\ 2015, 2016).  It  is generally agreed that DIB carriers probably are large carbon-based molecules (with $\sim$10-100 atoms) in the gas phase, such as polycyclic aromatic hydrocarbons (PAHs, e.g.\ Salama et al.\ 1996;  Salama 2008; Salama \& Ehrenfreund 2014), long carbon chains (e.g.\ Zach \& Maier 2014a,b) or fullerenes (e.g.\ Omont 2016).

In a recent paper (Omont, Bettinger \&  T\"{o}nshoff 2019, referred to hereafter as OBT19) we investigated more specific possible carriers among PAHs, addressing the case of 
polyacenes, C$_{\rm 4N+2}$H{$_{\rm 2N+4}$}, with N$\sim$10-18  fused rectilinear aligned hexagons. Polyacenes are attractive DIB carrier candidates because their high symmetry and large linear size allow them to form regular series of bands in the visible range with strengths exceeding that of most other PAHs, as confirmed by recent laboratory results up to undecacene (C$_{46}$H{$_{26}$}); see Shen et al.\ (2018) and references therein.

Polyacenes with very strong bands in the DIB spectral domain are just at the limit of stability against UV photodissociation.  
They are part of the prominent PAH family of interstellar carbon compounds, meaning that only $\sim$10$^{-5}$ of the total PAH abundance is enough to account for a medium-strength DIB.
So far, no definite inconsistency has been identified that would preclude polyacenes from being the carriers of some DIBs with medium or weak strength, including  the so-called C$_2$ DIBs. 
However, additional experimental data about the values of the gas-phase wavelengths of the optical bands of polyacenes are needed for firm conclusions.

In addition to their very strong optical bands and their high A rotation constant, which could fit the DIB rotation profiles, one of the main arguments supporting such elongated PAHs as DIB carriers is that they fit the general correlation that we found between the DIB wavelength and the apparent UV resilience of their carrier remarkably well (see Fig.\ 4 of OBT19; this is reproduced here in Fig.\ 5). 

This correlation  was established by studying the ratio R$_{21}$ of the equivalent widths of each DIB in two sightlines, HD\,183143 and HD\,204827, with different UV intensities. It is thought that  
R$_{21}$\,=\,EW(204827)/EW(183143) reflects the DIB carrier sensitivity to  destruction (or formation) by interstellar UV radiation. Interstellar lines of C$_2$ molecules have not been detected towards HD\,183143, while a very high value of N(C$_2$)/E(B-V) is observed towards HD\,204827, as well as C$_3$ and high column densities of diatomics (Thorburn et al.\ 2003; Hobbs et al.\ 2009, 2008). This is  
suggestive of relatively weak ambient radiation fields and/or shielded environments (see e.g.\ Welty 2014; Welty et al.\ 2014). 
The higher EW(5797)/EW(5780) DIB ratio values, such as are observed towards  HD\,204827, are  found only for lower inferred UV intensities 
(Welty 2014).
This $\lambda$/UV correlation could be  explained by the combined dependence on the wavelength of the strong optical bands of polyacenes and the UV photodissociation on their length.

The aim of this paper is to further explore this correlation for the different classes of DIBs. We more specifically address the dependence on  the DIB  band width because it was shown that the correlation is well obeyed by the prominent members of two main DIB families,  $\zeta$ and C$_2$, which have characteristic narrow widths. As pointed out in OBT19, various elongated molecules with $\pi$ electron conjugated systems should display a similar correlation. We here explore this possibility especially for the catacondensed PAHs 
and the long carbon chains. 

The paper is organised as follows. Section 2 recalls the width distribution of the DIBs and  proposes an optimal breakdown into two width classes that are compatible with the established DIB families. Section 3 analyses the observed $\lambda$/UV correlation for the two DIB classes by studying the wavelength dependence of the strength ratio  R$_{21}$ of each DIB in HD\,183143 and HD\,204827. 
This section establishes that the $\lambda$/UV correlation is dominated by  the narrow DIBs, but not the broader ones, including the $\sigma$ DIBs. 
The properties of  various $\pi$ electron conjugated systems are recalled in Section 4 in order to discuss the likelihood that they might account for the observed correlation. The possible role of the predominance of cationic PAH bands at large $\lambda$ is also considered, and it is shown that the infrared DIBs do not follow the general $\lambda$/UV correlation.  Finally, Section 5 summarises the possible amount of interstellar carbon that is locked up  in the different classes of DIBs depending on the values of their oscillator strengths. Section 6 connects the total DIB absorption and the amount of interstellar visible continuum absorption due to PAHs.

\section{Narrow and broad DIBs}

\begin{table}[htbp]
      \caption[]{Some identified DIB family members}
         \label{tab:lines}
            \begin{tabular}{ l l l }
            \hline
           \noalign{\smallskip}
 Family   &   DIBs(\AA ) & References   \\
      \noalign{\smallskip}
$\sigma$  & 4428  4502 4762  4885 5487   5705  & 1 \\
  & 5778  5780  5845  6010 6204 6283  &   \\ 
 & &   \\ 
$\zeta$  &5494 5545  5797  5850  6196  6376 &  2 \\     
 & 6379  6614 &   \\ 
 & &   \\ 
 C$_2$  &  4727 4969 4985 5176 5419 5513  & 3 \\
 &  5542 5546 5763 5793 6729 &   \\    
 &   \\ 
      \noalign{\smallskip}
            \hline
           \end{tabular}
{\small \begin{list}{}{} 
\item[$^1$ Lan et al.\ (2015) (Fig.\ 13; the 5845\,\AA\ DIB has been ]
\item[excluded because it is an ill-defined blend)] 
\item[$^2$ Ensor et al.\ (2017)] 
\item[$^3$ Thorburn et al.\ 2003,  DIBs detected in both HD\,183143]  
\item[and HD\,204827 (Hobbs et al.\ 2009, 2008) and confirmed by]
\item[Fan et al.\ (2019)]   
\end{list}}
\end{table}

A key property of the DIBs is their grouping into broad "families" whose behaviour depends on the interstellar environment and especially on the UV intensity (Krelowski \& Walker 1987; Herbig 1995). 
The strengths (equivalent widths) of the three main DIB families, $\zeta$, $\sigma,$ and C$_2$, are  correlated with the amount in the line of sight of H$_2$ 
(see Fig.\ 13 of Lan et al.\ 2015), HI, and C$_2$, respectively (e.g. 
 Herbig 1995; Cami et al.\ 1997; Vos et al.\ 2011; Ensor et al.\ 2017; Fan et al.\ 2017; Thorburn et al.\ 2003; Ka{\'z}mierczak et al.\ 2010; Elyajouri et al.\ 2018).


\begin{figure}[htbp]
         \begin{center}
\includegraphics[scale=0.7, angle=90]{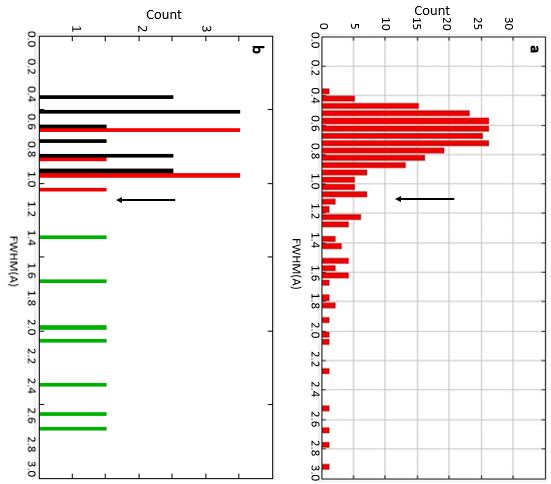}
 \caption{{\bf a)} Histogram of the values of the band width FWHM (\AA ) of the DIBs detected in the lines of sight of HD\,183143 or HD\,204827 (Hobbs et al.\ 2009, 2008) that have recently been confirmed in the Apache Point catalogue of Fan et al.\ (2019) for an FWHM\,$<$3\,\AA\ (out of 509 DIBs, there are 19 with FWHM\,$>$3\,\AA ). {\bf b)} Histogram of the values of the band width of the DIBs, identified as C$_2$ (black), $\zeta$ (red), and $\sigma$ (green) (Table 1), detected in the lines of sight of HD\,183143 or HD\,204827, for FWHM\,$<$3\,\AA\ (out of 12 $\sigma$ DIBs,  there are 4 with FWHM\,$>$3\,\AA ).
The vertical arrows show the adopted limit, 1.1\,\AA , for the separation between "narrow" and "broad " DIBs.}
     \end{center}
 \end{figure}

The general correlation apparent in Fig.\ 4 of OBT19 (reproduced in Fig.\ 5) between the UV sensitivity and the wavelength of the DIBs mixes several DIB families, which each have a different dependence on the UV intensity. One obvious way to learn more about the origin of this correlation is to determine how it specifically affects the main DIB families. The correlation might be more pronounced for the broad class of  narrow DIBs  that includes the C$_2$  and $\zeta$ DIBs,  which both display similar band profiles (e.g.\ Elyajouri et al.\ 2018). 

However, the actual classification of most DIBs into such families remains difficult because their correlations are complex and their spectra are noisy. In addition to the specific case of the C$_2$ DIBs, no more than 12 strong DIBs have a well-established membership in each of the $\zeta$ and $\sigma$ families (Table 1). In order to better understand the $\lambda$-UV correlation, it is therefore much simpler to address its dependence on the DIB width, which is relatively easy to measure for all DIBs (e.g.\ Fan et al.\ 2019) and is related to the families. 
It is known that most of the $\zeta$ DIBs display a structured narrow profile (e.g.\ Herbig 1995; Sarre 2014 and references therein), similar to that of the C$_2$ DIBs (Elyajouri et al.\ 2018), while  the $\sigma$ DIBs have a broader profile (Fig.\ 1b).


\begin{figure}[htbp]
         \begin{center}
\includegraphics[scale=0.65, angle=0]{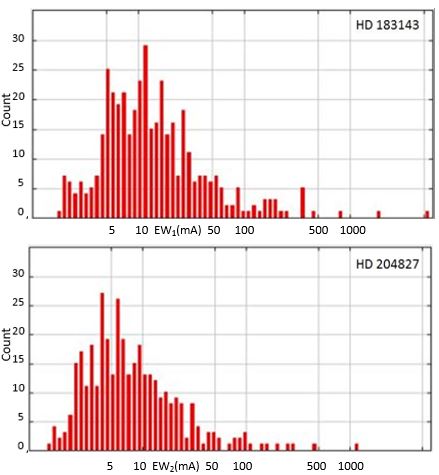}
 \caption{Histograms of the equivalent widths EW1 (top panel) and EW2 (lower panel)  (m\AA ) of the DIBs detected in the lines of sight of HD\,183143 and HD\,204827, respectively, from Hobbs et al.\ (2009, 2008), and confirmed by Fan et al.\ (2019).}
     \end{center}
 \end{figure}

\begin{table}[htbp]
      \caption[]{Properties of narrow and broad DIBs detected in the lines of sight of HD\,183143 (1) and HD\,204827 (2),  derived when adopting a limit of 1.1\,\AA\ between both classes}
         \label{tab:lines}
            \begin{tabular}{ l c c  c c    }
            \hline 
           \noalign{\smallskip}
 {\tiny DIBs detected in} &   {\tiny 1 and 2$^a$}  &{\tiny 1  not 2$^b$}  &  {\tiny  2  not 1$^c$} & {\tiny Total} \\
      \noalign{\smallskip}
  Narrow  {\tiny($<$1.1\AA )}  &  &   & &  \\       
{\small Number}   &  218  &  83 &  85  & 386  \\
EW$_{\rm 1av}$ & 19.4  &   9.8  &  x  &  x     \\  
EW$_{\rm 1tot}^d$ & 4230 &   810  &  x  & 5040 \\  
EW$_{\rm 2av}$ & 12.0  & x &   6.3     &  x    \\  
EW$_{\rm 2tot}^d$ & 2620 & x&   540     & 3160 \\  
          \hline
  & & & &  \\
Broad$^e$ {\tiny($>$1.1\AA )} & & & &  \\       
{\small Number}  & 49   &  67 & 6    & 122  \\
EW$_{\rm 1av}$ & 133 &   38  & x   &   x     \\  
EW$_{\rm 1tot}^d$ & 6530 &   2550  & x &9080 \\  
EW$_{\rm 2av}$ & 47     & x  &  26.6     &   x     \\  
EW$_{\rm 2tot}^d$ &2310  &x & 160     & 2470  \\ 
          \hline
  & & & &  \\
4428\,\AA & & & & \\
EW$_1$ & 5700 & x &x & 5700 \\
EW$_2$ & 1220 & x &x & 1220  \\
      \noalign{\smallskip}
            \noalign{\smallskip}
            \hline
           \end{tabular}
{\small \begin{list}{}{} 
\item[]   $^a$   Confirmed DIBs detected in both lines of site (Hobbs et al.\ 2008, 2009; Fan et al.\ 2019)
\item[]   $^b$  detected only in HD\,183143 (1)
\item[]   $^c$  detected only in HD\,204827 (2)
\item[]   $^d$  Sum of the equivalent widths of all DIBs of the quoted class detected in the quoted sightlines (see Section 5)
\item[]  $^e$  Not including the 4428\,\AA\ DIB
\item[]  $^f$  All  equivalent widths, EW$_i$, have m\AA\ units
\end{list}}
\end{table}

\begin{figure}[htbp]
         \begin{center}
\includegraphics[scale=0.65, angle=0]{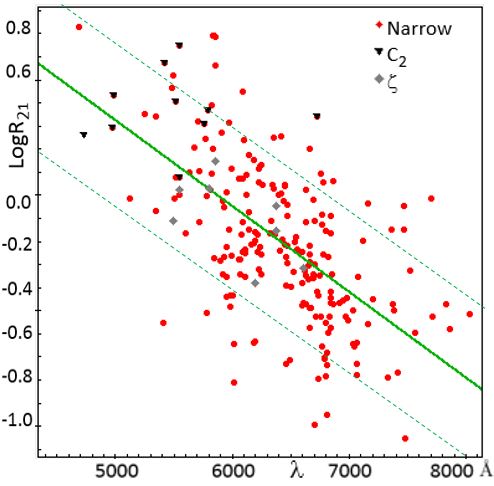}
 \caption{Wavelength dependence of the ratio R$_{21}$ of the equivalent widths of narrow DIBs (FWHM\,$<$\,1.1\,\AA , Table 2) in the lines of sight of  HD\,204827 and HD\,183143 (Hobbs et al.\ 2008, 2009)  confirmed in Fan et al.\ (2019). These DIBs are shown by red dots, except for the DIBs identified in Table 1 as $\zeta$ (Ensor et al.\ 2017; grey diamonds) or C$_2$ (Thorburn et al.\ 2003; black triangles). The straight full green line displays the linear fit of Eq(1), and the dashed lines show the $\pm$1$\sigma$ limits.} 
     \end{center}
 \end{figure}

\begin{figure}[htbp]
         \begin{center}
\includegraphics[scale=0.63, angle=0]{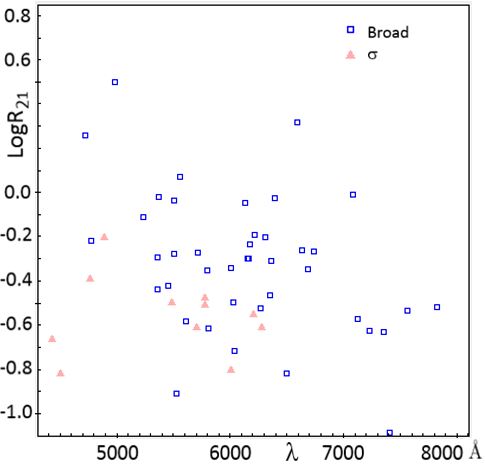}
 \caption{Same as Fig.\ 3 for broad DIBs (FWHM\,$>$\,1.1\,\AA , Table 2). These DIBs are shown by blue open squares, except for the DIBs identified in Table 1 as $\sigma$ (Lan et al.\ 2015) which appear as orange full triangles. 
 }
     \end{center}
 \end{figure}

\begin{figure}[htbp]
         \begin{center}
\includegraphics[scale=0.61, angle=0]{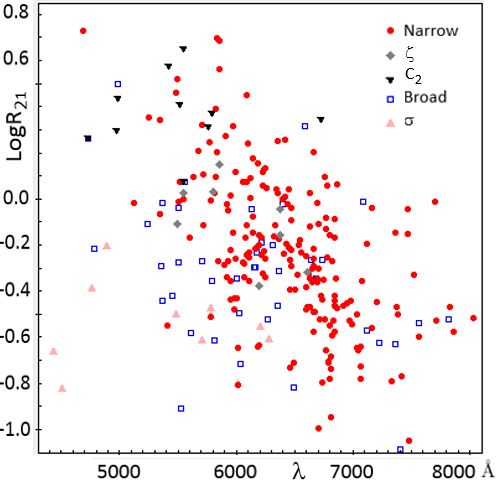}
 \caption{Same as Fig.\ 3 for all DIBs (same as Fig.\ 4 of OBT19),  with the same symbols as in Figs.\ 3 and 4.
 }
     \end{center}
 \end{figure}

 The histogram of the bandwidth full width at half-maximum (FWHM), taken from  the Apache Point catalogue of Fan et al.\ (2019), of the 509 DIBs  detected in the sightlines of HD\,183143 (number 1) or HD\,204827 (number 2) by Hobbs et al.\ (2009, 2008) and confirmed in Fan et al.\ (2019), is displayed in Fig.\ 1a, while the histograms of their equivalent widths, EW1 and EW2, respectively, are shown in Fig.\ 2. The FWHM histogram of Fig.\ 1a shows an accumulation of narrow DIBs with FWHM\,$\la$\,1\,\AA . However, there is no clear-cut sharp transition in the histogram between narrow and broad DIBs. In Table 2 we therefore chose a somewhat arbitrary definition of "narrow" DIBs,  that is, an\ FWHM\,$<$\,1.1\,\AA , because the histogram (Fig.\ 1a) shows a shallow minimum for this value.
 Other definitions are possible, with slightly different values of the limit that separates narrow and broad DIBs, such as 1.0\,\AA\  (see Appendix A).
As shown in Figs.\ 1b and 3,  all $\zeta$ and C$_2$ DIBs listed in Table 1 are included in the region of the narrow DIBs with an FWHM\,$<$\,1.1\,\AA , while those listed as $\sigma$ are outside and much broader for most of them.  We verified that the situation would not have been very different for  other definitions of narrow DIBs. 

  Some  properties of this set of 386 narrow DIBs (76\% of the total, 509) are reported in Table 2 and discussed in Section 5, together with the complementary set  (123, 24\%) that we call "broad" DIBs. 
With any  possible definition of narrow DIBs 
(e.g.\ Tables 2 and A.1),
their number  is very high, from $\sim$70\% to $\sim$80\% of all DIBs. However, they contribute only a minor fraction to the total strength (equivalent width) of the DIBs:  $\sim$20\%-27\% in the line of sight of HD\,183143; $\sim$40\%-47\% in HD\,204827, where the narrow $\zeta$ and C$_2$ DIBs are enhanced (see Tables 2 and A.1 and the discussion in Section 5).

\section{Correlation between the DIB UV sensitivity and their wavelength}

\subsection{DIBs detected in the lines of sight of  HD\,204827 and HD\,183143}

With this definition of narrow and broad DIBs (Table 2), Figs.\ 3 and 4 break down the general correlation shown in Figure 4 of OBT19 (reproduced in Fig.\ 5) 
between the DIB intensity ratio, R$_{21}$, in the sightlines of  HD\,204827 and HD\,183143 for narrow and broad DIBs separately. Because the UV intensity is attenuated in the sightline of HD\,204827 but retains the typical high value of diffuse clouds in  HD\,183143, R$_{21}$ reflects the DIB carrier sensitivity to its destruction (or formation) by interstellar UV. Figure 3 shows that the correlation is significantly higher for narrow DIBs than for all DIBs (Fig.\ 5). A linear fit gives

\begin{equation}
 {\rm logR_{21}\,=\, a\,\times\,10^{-4}\times\lambda(A)\,+\,b\,\pm\,c}
\end{equation}

 with a\,=\,-3.71, b\,=\,2.18, and c\,=\,0.28, with a Pearson correlation coefficient  P\,=\,-0.61 for the narrow DIBs (Fig.\ 3), instead of  a\,=\,-2.4, b\,=\,1.3, c\,=\,0.32,  and P\,=\,-0.45 for all DIBs (Fig.\ 5). On the other hand, hardly any correlation is visible  for the majority of broad DIBs (Fig.\ 4) (a\,=\,-0.85, b\,=\,0.15, c\,=\,0.33,  and P\,=\,-0.22).
These coefficients confirm the significant correlation for the narrow DIBs, but with a substantial dispersion of the values of R$_{21}$ around this fit by a factor of about 2 (Fig.\ 3). The large uncertainties on equivalent-width measurements, which are particularly significant for the numerous weaker DIBs ($<$10m\AA ), is expected to substantially contribute to this dispersion. However, their effect on the correlation itself is apparently negligible.

We verified that similar $\lambda$/UV correlations are obtained for other possible definitions of narrow DIBs, but with a slightly less clear-cut difference between narrow and broad DIBs. For instance, the case of a separation between narrow and broad DIBs of 1.0\,\AA\  is discussed in Appendix A and displayed in  Fig.\ A.1. The fit and Pearson coefficients are almost identical to those of Fig.\ 3 (see Eq.\ B.1). 
We may conclude that the restriction to narrow DIBs significantly enhances the $\lambda$/UV correlation, while the broadest DIBs  do not display such a correlation. However, it is not possible to provide a completely clear-cut definition of narrow DIBs.

\subsection{DIBs detected on a single sightline}


\begin{figure}[htbp]
         \begin{center}
\includegraphics[scale=0.63, angle=0]{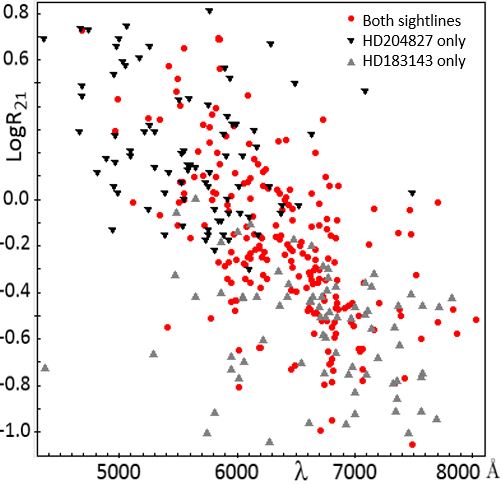}
 \caption{Wavelength dependence of the ratio R$_{21}$ of the equivalent widths of narrow DIBs (Table 2) in the lines of sight of  HD\,204827 and HD\,183143 (Hobbs et al.\ 2008, 2009; Fan et al.\ 2019), replacing the equivalent widths of undetected DIBs by 3\,m\AA\ in HD\,183143 and 2\,m\AA\ in HD\,204827.}
     \end{center}
 \end{figure}


\begin{figure}[htbp]
         \begin{center}
\includegraphics[scale=0.63, angle=0]{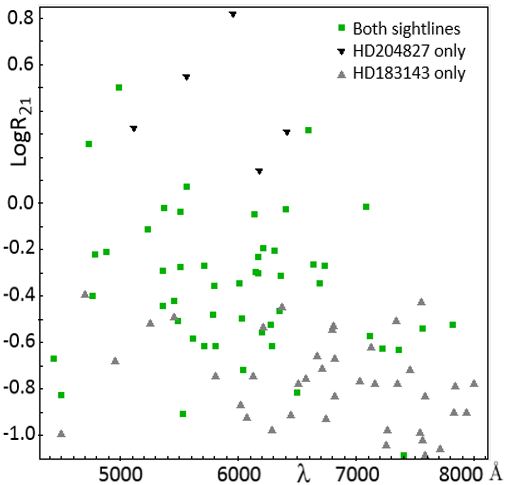}
 \caption{Same as Fig.\ 6, but for broad DIBs.}
     \end{center}
 \end{figure}

The numerous DIBs detected on only a single sightline, either  HD\,183143 (Hobbs et al.\ 2009) or HD\,204827 (Hobbs et al.\ 2008), may provide additional information about the $\lambda$/UV correlation. As quoted in OBT19,  168 DIBs are tentatively detected only in HD\,183143 and 107 in HD\,204827 as reported by Hobbs et al.\ (2009, 2008). We here retain only those confirmed in the recent Apache Point catalogue of Fan et al.\ 2019, that is,\ 150 in HD\,183143 and 91 in HD\,204827. Their overwhelming presence at short wavelength in HD\,204827  (89\% at $\lambda$\,$<$\,6200\,\AA ) and at long wavelength in HD\,183143 (65\%\ at $\lambda$\,$>$\,6500\,\AA ) additionally supports the correlation of the ratio R$_{21}$ with $\lambda$ and therefore the  $\lambda$/UV correlation. 

This point may be illustrated by extending Figure 3 to DIBs that are detected on a single sightline. For this purpose, we replace in Figure 6 the equivalent widths of undetected DIBs by the somewhat arbitrary values of 3\,m\AA\ in HD\,183143 and 2\,m\AA\ in HD\,204827, which can be consistent with the low-intensity parts of the histograms of the equivalent widths in Figure 2. For narrow and broad DIBs, the agreement with the behaviour of the DIBs detected in the two lines of sight is remarkable. It remains somewhat artificial because it depends on the adopted choice for the intensity of undetected DIBs. Nevertheless, Figure 6 shows that the $\lambda$/UV correlation extends from the 218  narrow DIBs detected in both sightlines to almost 400 detected in at least one of them (the fit and Pearson coefficients are a\,=\,-3.88, b\,=\,2.2, c\,=\,0.44,  and P\,=\,-0.66). Similarly, Figure 7 shows that there is no strong correlation for the 123 broad DIBs  that are detected in at least one sightline (a\,=\,-2.1, b\,=\,0.64, c\,=\,0.45, and P\,=\,-0.33).

Several tens of DIBs are now detected outside the range of the catalogues of Fan et al.\ (2019) and Hobbs et al.\ (2008, 2009),  in the near-infrared from 0.8 to 2.4\,$\mu$m (Joblin et al.\ (1990);  Geballe et al.\ 2011, 2014; Cox et al.\ 2014; Elyajouri et al.\ 2016, 2017; Hamano et al.\ 2016, 2018; Galazutdinov et al.\ 2017), including the C$_{60}^+$ DIBs near 9600\,\AA \ and the 8620\,\AA \ "Gaia DIB"  (e.g.\ Kos et al.\ 2013; Puspitarini et al.\ 2015; Krelowski et al.\ 2019). Only a  few of them have been observed in  highly UV-screened lines of sight, such as  HD\,204827, with the required sensitivity that the $\lambda$/UV correlation needs to be determined. However, as discussed in Section 4.6, a tight correlation is not expected because the majority of them seems to behave like cationic $\sigma$ DIBs.

\section{Implications for carriers of narrow DIBs}

\subsection{Large neutral elongated carbonaceous molecules}

As quoted in OBT19,  the series of strong optical bands of families of carbonaceous elongated molecules might fit the correlation of R$_{21}$  with $\lambda$ well under the condition that their carriers are critically sensitive to interstellar UV photodissociation. The remarkable optical properties of this essential class of organic molecules with conjugated $\pi$ electron systems, including the linear progression of $\lambda$ with the number of carbon atoms N$_{\rm C}$ and f\,$\ga$\,1,  have been rationalised a long time ago (e.g.\ Dewar \& Longuet-Higgins 1954; Platt 1956; and the papers of  Brooker et al.\ 1940-1951 quoted by Platt 1956). They include all varieties of carbon chains and catacondensed PAHs and their derived compounds, and more generally, most of the organic dyes. This series of strong optical bands can be traced back to the simple free electron model (e.g.\ Jensen 2013 and historical references therein), but the account of actual values of the band wavelengths and strengths requires various levels of sophistication of quantum theory. The interest for DIB carriers of some classes of these compounds have been  emphasised for more than 40 years for long carbon chains (Douglas 1977),  more recently for catacondensed PAHs (Ruiterkamp  et al.\ 2002, 2005), and particularly polyacenes (OBT19). In the following subsections, we summarise the conclusions for each class and try to re-examine them in the context of the $\lambda$/UV behaviour of narrow DIBs. 

A key interest of compounds such as DIB carriers is the rough proportionality of the wavelength of their strong optical bands with their length. Their interstellar UV  photodissociation is also strongly correlated with their length because after the absorption of a UV photon, their vibration temperature T$_{\rm v}$ is roughly inversely proportional to N$_{\rm C}^{0.4}$ (Eq.\ B.3) and  similarly  to their length. Because the probability of photodissociation exponentially varies with T$_{\rm v}$, the sensitivity to UV photodissociation must be correlated with the length and thus with $\lambda$. 

As discussed in OBT19, the dehydrogenation of  N-acenes and similar molecules depends very strongly on N around a critical value N$_{\rm cr}$.  As stressed in Appendix B, N$_{\rm cr}$ also depends critically on the effective energy E$_{\rm UV}$ of the photodissociating photons. This may explain the order of magnitude of the observed R$_{21}$/$\lambda$ correlation. In a substantial part of the sightline of HD\,204827, the UV-shielding is high enough to preserve the hydrogenation down to N\,$\sim$\,11, while it is destroyed in HD\,183143. This may explain the high values of R$_{21}$ that are observed in the range $\lambda$\,$\sim$\,5000\,\AA . On the other hand, 
all acene-like molecules are expected to be fully hydrogenated for N\,$\ga$\,15 ($\lambda\,\ga$\,6500\,\AA ), yielding relatively low values of R$_{21}$.

However,  other origins for the correlation of R$_{21}$  with $\lambda$ are possible.  It is likely that a similar correlation  exists, although less marked, for pericondensed PAHs between the wavelength of  the optical bands and N$_{\rm C}$ and hence the probability of photodissociation. This is in line with the well-known propensity of all neutral PAHs to have narrow bands (e.g.\ Salama 2008). Alternatively, as quoted in OBT19, the correlation might also partly result from the predominance of cationic bands at large $\lambda$ (Section 4.3).

\subsection{Carbon chains}

The interest of  long  carbon chains as DIB carriers has been emphasised for more than 40 years (Douglas 1977, Thaddeus 1995, and especially John Maier's Basel group, e.g.\  Rice \& Maier 1993; Fischer \& Maier 1997; Zack \& Maier 2014a,b; Campbell \& Maier 2017). It is generally agreed, however, that chains with N$_{\rm C}$\,$\la$\,15 are excluded based on laboratory results (e.g.\ Zack \& Maier 2014a) probably because they are efficiently destroyed by single-photon photodissociation. 

For longer chains, the N$_{\rm C}$ stability limit  in the diffuse interstellar medium (ISM) might be inferred from an analogy with PAHs. Modelling PAH photodissociation remains difficult (e.g.\ Simon et al.\ 2018 and references summarised in OBT19). It is generally agreed that dehydrogenation occurs first for N$_{\rm C}$\,$\la$\,40-50, and that the carbon skeleton is efficiently destroyed in typical diffuse clouds for slightly lower values of N$_{\rm C}$\,$\la$\,30-40 (e.g.\ Allain et al.\ 1996a,b; 
Castellanos et al.\ 2018a,b and references therein). The case of long chains is expected to be relatively similar, with similar  critical values,  N$_{\rm Ccr}$, of  N$_{\rm C}$  because they have roughly similar dissociation energies for the ejection of H and carbon clusters  (mostly C$_3$ for carbon chains, Chabot et al.\ 2010). 

In typical diffuse clouds, chains with N$_{\rm C}$\,$\ga$\,30-40
might perhaps survive and grow by C$^+$ accretion. They might eventually bear an hetero atom  (N, etc.) at their ends, and even a few H atoms  for slightly higher values of  N$_{\rm C}$. If their photodissociation worked like this, it might 
explain  a $\lambda$/UV correlation in a similar way as for polyacenes if they display series of strong optical bands and N$_{\rm Ccr}$ lies in the visible  range.  
The spectra of carbon rings present similarities with chains, and rings might spontaneously form from very long chains in the conditions of the ISM.

However, because smaller chains are expected to be destroyed by photolysis, it is generally believed that  chain 'seeds' would be needed to allow for  free very long chains. 
There is still no evidence of such 
chains  in the diffuse interstellar medium, however, and the proposed processes
for their formation from grain shattering (e.g. Duley 2000; Jones
2016) remain tentative.
In addition, the properties of DIB profiles seem incompatible with such long linear chains (Oka et al.\ 2013, Huang et al.\ 2013).
Nevertheless, we note that a suite of alkyl dicarboxylic acids up to C$_{18}$ chain length is observed in some  meteorites (Pizzarello \& Shock 2010).

\subsection{Polyacenes}

Polyacenes remain the best candidates to contribute to the DIB $\lambda$/UV correlation based on the arguments presented in OBT19 especially for narrow  C$_2$ and $\zeta$ DIBs, namely their high symmetry and  series of very strong optical bands, combined with the properties of their photodissociation, may fit  the $\lambda$/UV correlation remarkably well; their large A rotational constant may account for the observed narrow profiles of C$_2$ and $\zeta$ DIBs; and they are members of the extremely abundant PAH family. 

In addition to fully hydrogenated polyacenes, partially dehydrogenated radical states also need to be considered for N$_{\rm C}$ close to the critical value for photodissociation N$_{\rm Ccr}$, as discussed in OBT19. However, a small number of singly dehydrogenated acenes might also be created for N$_{\rm C}$\,$>$\,N$_{\rm Ccr}$ from photodissociation of transient  superhydrogenated states producing H$_2$ (Boschman et al.\ 2015; Castellanos et al.\ 2018a,b; Sanchez-Sanchez et al.\ 2019). 

If pure acenes were proven to be carriers of strong or medium DIBs, the rich variety of their possible alterations as possible carriers of weak DIBs need to be considered as well. Many of them are expected to keep the most  characteristic band features of acenes into wavelength series and strengths, although predicting precise values is difficult. This could include 
 superhydrogenation (with several H atoms bound to the same edge C atom), methylation, C addition, or inclusion of hetero atoms (N, O, etc., e.g.\ Hudgins et al.\ 2005; Tielens 2008). 

However, such modifications must be compatible with polyacene interstellar chemistry  and the probability that  they can survive  in the harsh UV environment of the diffuse ISM to warrant significant abundances.
In contrast, the activation energy of many adducts is not high enough to protect them against photolysis by single UV-photon absorption in the diffuse ISM. This is the case for the H atoms of superhydrogenated states (Castellanos et al.\ 2018a,b) and for CO ejection from quinones (Chen et al.\ 2018) in PAHs with N$_{\rm C}$\,$\la$\,50-70.
In addition,  in most cases, including nitrogen atom substitutions in the
carbon skeleton, part of the high symmetry of polyacenes is expected to be broken, and the actual intensity of possible interstellar bands is expected to be reduced by their isomer variety.

As quoted in OBT19, the lower thermodynamic stability of linear polyacenes compared to compact PAHs remains a major question for their abundance in the diffuse ISM. While only $\sim$10$^{-5}$ of the total PAH abundance is enough to account for a medium-strength DIB (or $\sim$0.3-0.4\% of the PAH abundance for the combined strengths of all narrow DIBs, see Section 5), some processes allowing such an  abundance would need to be identified. For instance, the possibility of reactions on grain surfaces leading to the  formation of elongated PAHs perhaps through radical and polyacene reactions would need to be identified  (e.g.\ Johansson et al.\ 2018; Sanchez-Sanchez et al.\ 2019; Zuzak et al.\ 2018). One of the most efficient modes of synthesis of graphene nanoribbons (GNR) similarly uses polymer formation on surfaces (e.g.\
Cai et al.\ 2010; Ruffieux et al.\ 2016; Nakada et al.\ 1996).  Even oxygen unzip of graphenic interstellar sheets might perhaps be considered, as it is known for laboratory graphene or nanotubes  (Li et al.\ 2006, Ajayan \& Yakobson 2006; Kosynkin et al.\ 2009). Because polyacenes and other catacondensed PAHs discussed below are extreme forms of GNRs, it is not excluded that they from similarly efficiently by carbon chemistry on the surface of interstellar (carbon) grains. The interstellar (photo-)chemistry of dust grains and PAHs is too complex and still too poorly understood to allow any definitive conclusion, however. 

\subsection{Other catacondensed PAHs}

\begin{figure}[htbp]
         \begin{center}
\includegraphics[scale=0.62, angle=0]{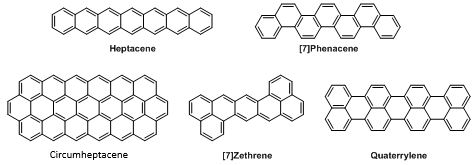}
 \caption{Examples of elongated PAHs from five different classes: acenes, phenacenes, circumacenes, zethrenes, and rylenes.}
     \end{center}
 \end{figure}

While acenes are the most extreme and symmetric cases, various other elongated or catacondensed PAHs (see e.g.\ Fig.\ 8) share many of the polyacene properties. 
Their  conjugated $\pi$ electrons are expected to also yield similar series of strong bands. 
It is possible that the broader acene family and other polyacene  based isomers, such as  zethrenes, fit the $\lambda$/UV correlation as well as polyacenes because their optical bands and photodissociation versus length share similar properties. 
Oligorylenes  are especially attractive candidates (Ruiterkamp et al.\ 2002, 2005;  Halasinski et al.\ 2003; Zeng et al.\ 2018), 
as are circumacenes (Malloci et al.\ 2011), because their highest occupied molecular orbital - lowest unoccupied molecular orbital (HOMO-LUMO) 
gap is small and the f-values of their optical bands are high (Weisman et al.\ 2003; Zdetsis \& Economou  2017). 
They have other properties similar to acenes, such as diradical structure, but they are expected to be slightly more stable and might intervene in carbon graphitisation (see e.g.\ the review by Zeng et al.\ 2018 and references therein). 
The discussion of Ruiterkamp et al.\  (2005) shows that the measured bands of neutral and cation terylene (C$_{30}$H$_{16}$) and quaterrylene (C$_{40}$H$_{20}$) in a neon matrix (Ruiterkamp et al.\  2002) seem incompatible with strong DIBs. This might be explained by photodissociation, so that heavier rylenes might nevertheless be 
considered for strong $\zeta$ DIBs with $\lambda$\,$\ga$\,6500\,\AA\ and IR $\sigma$ DIBs.
On the other hand, compounds such as phenacenes appear less favourable because their  HOMO-LUMO gap is larger (e.g.\ Malloci et al.\ 2011).

Together with their linear form,  the cyclic form of acenes, [N]cyclacenes, C$_{\rm 4N}$H{$_{\rm 2N}$}, might be present in the interstellar medium (Jones 2016; OBT19).  
Theoretical studies of their electronic properties (Wu et al.\ 2016, Battaglia et al.\ 2017) have confirmed that the properties of cyclacenes with an even number of benzene rings are similar to linear acenes. We infer that they probably display similar visible  spectra of conjugated $\pi$ electrons, but the wavelengths and strengths remain to be determined.    

Cyclacenes could efficiently form in the diffuse ISM from long linear radical acenes (Sect.\ 4.3) through monomolecular cyclisation reactions. The frequent absorption of a $\sim$10\,eV UV photon could raise the internal temperature to overcome the few eV bending barrier to achieve cyclisation through a monomolecular reaction similar to the radical or superhydrogenation reactions between  polyacenes  studied by Johansson et al.\ (2018) or Sanchez-Sanchez et al.\  (2019).

\subsection{Cations}
 As quoted in OBT19, the correlation might also result partly from the predominance of cationic bands at large $\lambda$. Cations are expected to dominate the visible  absorption of interstellar PAHs (e.g.\ Salama et al.\ 1996; Li \& Draine 2001 and Section 6).  It is expected that they display many more bands in the range $\sim$1-2\,eV than  their parent closed-shell neutrals because an electron might be promoted from the inner shells into the vacancy of the HOMO shell (see e.g.\ Salama et al.\ 1996). However, the actual spectral distribution of their expected interstellar bands remains uncertain because it depends on the nature of the most important PAHs. This may at least partly explain the large number of DIBs with a low R$_{21}$ ratio or those that are undetected on the sightline of HD\,204827 at $\lambda$\,$\ga$\,6000\,\AA\ (Figs.\  3 \& 6).  

Heavy ionised acenes have relatively strong optical and infrared bands, although they remain weaker than the optical bands  of neutral polyacenes. They have been measured up to nonacene by Mondal et al.\ (2009) and T{\"o}nshoff \& Bettinger (2010) (see also Malloci et al.\ 2007). Larger acenes also have lower ionisation potentials. This is expected to further push the charge state balance towards cations and increase the number  of cationic DIBs at $\lambda$\,$>$\,6000\,\AA .

\subsection{Infrared DIBs}
As shown for the C$_{60}^+$ DIBs near 9600\,\AA , similar cationic bands are expected for all fullerenes and various PAHs in the whole near-infrared range beyond 8000\,\AA . The detection of more than 60 near-infrared DIBs has been announced by various authors since the first detections (Joblin et al.\ 1990; Geballe et al.\ 2011, 2014), (see Section 3.2 and below).
However, the data and even the list are not yet public for many of those that were  announced by Hamano et al.\ (2018). For the others, fewer than 12, the strongest, have published  observations that are complete enough to precisely determine their width and their dependence on the UV environment. This includes the following DIBs: $\lambda \lambda$10780, 10792, 11797, 12623,  13175  (Hamano et al.\ 2016) and $\lambda$15268\footnote {We give here the value in the air of the wavelength of this DIB to be consistent with the other DIBs, while Elyajouri et al.\ (2017) used its value in vacuum, $\lambda$15273.} (Elyajouri et al.\ 2017), in addition to the $\lambda \lambda$9577 and 9633 C$_{60}^+$ DIBs (Foing \& Ehrenfreund 1997; Cox et al.\ 2014) and the $\lambda$8620 "Gaia DIB" (Krelowski et al.\ 2019). All of them 
probably form in similar environmental conditions as the optical DIBs at $\lambda$5780 or even $\lambda$6283, that is, in a relatively strong UV radiation field, as quoted by Elyajouri et al.\ (2017), for example. They may therefore be considered as $\sigma$ DIBs and their carriers are probably cations, as confirmed for the $\lambda$9577 and 9633 C$_{60}^+$ bands. Although only $\lambda$8620 and $\lambda$15268 seem to have published data in HD\,204827 (Krelowski et al.\ 2019; Galazutdinov et al.\ 2017), we expect that the intensity ratio, R$_{21}$, on the sightlines of HD\,204827 and HD\,183143 of these 9 DIBs is comparable to those of  $\lambda$5780 (0.33), $\lambda$6283 (0.24), $\lambda$8620 (0.30), or $\lambda$15268 (0.20). This clearly places them well outside the range of the $\lambda$/R$_{21}$ correlation of Fig.\ 3. This is another example that the $\sigma$ DIBs do not follow the $\lambda$/UV correlation.

While most prominent near-infrared DIBs are $\sigma$-type, no $\zeta$-type has been well identified yet. 
The  possible different case of $\lambda$10504  is discussed  by Hamano et al.\ (2016).  However, the precise determination of the properties of this DIB is made difficult by the presence of stellar features in the spectra of the targets and in those of the comparison stars (R.\ Lallement private communication).

When the FWHM values of the detected  infrared DIBs  (Geballe et al.\ 2011; Cox et al.\ 2014; Hamano et al.\ 2015; Elyajouri et al.\ 2017;  Galazutdinov et al.\ 2017; Krelowski et al.\ 2019) are expressed in \AA ngstr\"{o}ms, they are all above the 1.1\,\AA\ limit that separates narrow from broad optical DIBs. However, expressing the FWHM in cm$^{-1}$ (or in km/s) is certainly more relevant when infrared and optical DIBs are compared. When the general relation 
\begin{equation}
{\rm FWHM(cm^{-1}) =  FWHM(A)/\lambda (\mu m)^2}
\end{equation}
is taken into account, we obtain an FWHM $\approx$ 3.0\,cm$^{-1}$ at about 6000\,\AA\  for the narrow or broad limit, FWHM\,=\,1.1\,\AA , that we assumed. Adopting 3.0\,cm$^{-1}$ for the limit regardless of $\lambda$ would  yield an FWHM(\AA )\,=\, 3$\times$$\lambda(\mu {\rm m})^2$ for this limit. We verified that for the optical DIBs, the R$_{21}$/$\lambda$ correlation and the values of the coefficients for the fit of Eq(1) are practically unchanged when a 3.0\,cm$^{-1}$ limit is adopted. However, most FWHM values of the infrared DIBs are now close to or below this limit value, including all the \textquotesingle $\sigma$\textquotesingle\  infrared  DIBs discussed above, except for $\lambda$8620. However, the relevance of this limit may be debated. We may conclude that while most prominent near-infrared DIBs are of $\sigma$-type, the $\lambda$/UV correlation is not relevant for infrared DIBs.

The density of DIBs, especially of narrow ones, sharply decreases above 7000\,\AA , and it decreases still more above 8000\,\AA , even after correcting the table of Fan et al.\ (2019) for telluric absorption. 
This noticeable lack of narrow DIBs above $\sim$8000\,\AA\ (which needs to be confirmed by the analysis of new infrared DIBs, e.g.\ Hamano et al.\ 2018) might imply that there is a limit to the size (number of rings) of the carriers that give rise to the $\lambda$/UV relation (e.g.\ for neutral polyacenes, Eq.\ (B.2) gives a maximum of about 19 rings for a 8000\,\AA\ limit). The origin of this size limit might be either a drop in abundance of the carriers or a change in their optical band strength following a structural change of their ground state.

\section{Amount of carbon implied in the different DIB classes}

It is straightforward to derive  the amount of carbon that is locked up in the carriers of a DIB or an ensemble of DIBs of known equivalent width and oscillator strength (e.g.\ Ruiterkamp et al.\ 2005; Cami 2014, Tielens 2014, Omont 2016). For instance, Eq(1) of OBT19 yields 

\begin{equation}
{\rm X_{CM} \approx 10^{-7} x~EW / (r~x~\bar{f}) }
,\end{equation}

where X$_{{\rm CM}}$ denotes the fraction of total interstellar carbon that is locked up in  
molecule(s) M assuming  N$_{\rm C}$\,$\sim$\,60 carbon atoms in M and ignoring the $\lambda$ dependence. $\bar{f}$ is some average of the DIB oscillator strength, EW is its equivalent width in m\AA , and r\,=\,E(B - V)  is the sightline reddening. 

Table 2 provides a rough statistics of the DIBs detected in the lines of sight of HD\,183143 and HD\,204827 (Hobbs et al.\ 2009, 2008; Fan et al.\ 2019) together with their average and total equivalent widths EW$_1$ and EW$_2$. From these data it is straightforward to derive the amount of carbon that is contained in the carriers of each DIB  class modulo the mean oscillator strength. 
For this purpose,  with the actual values of r\,=\,1.27\,mag  and 1.1\,mag,  
Eq.\ (1) yields 

\begin{equation}
{\rm X_{CM1} \approx 0.8~x~10^{-7} x~EW_1 /  \bar{f} }
\end{equation} 

\begin{equation}
{\rm X_{CM2} \approx 0.9~x~10^{-7} x~EW_2 / \bar{f} }
\end{equation}

for HD\,183143 and HD\,204827, respectively.
 From the values of Table 2, we derive the following  orders of magnitude of the total percentages  of interstellar carbon that is locked up in the different DIB classes  (these percentages are slightly but significantly modified by $\sim$10-20\% when 1.0\,\AA\ is assumed for the limit between narrow and broad DIBs, Table A.1).

$\bullet$ Narrow DIBs  ($<$\,1.1\,\AA ): The 386 narrow DIBs (76\% of all 509 DIBs) account for 25\% of the total equivalent width in HD\,183143) (46\% in HD\,204827) and 36\% (56\%) without the 4428\,\AA\ DIB. This corresponds to ${\rm 4~x~10^{-4} /  \bar{f} }$ of interstellar carbon in HD\,183143 and ${\rm 2.8~x~10^{-4} /  \bar{f} }$ in HD\,204827.
If their carriers are elongated PAHs with high f-values $\sim$1, they contain only a very small fraction of the interstellar carbon, about 0.3-0.4\% of that in all PAHs.

$\bullet$  Broad DIBs  ($>$\,1.1\,\AA ): The 122 broad DIBs excluding the 4428\,\AA\ DIB (24\% of all DIBs) account for 46\% of the total equivalent width in HD\,183143 (36\% in HD\,204827) and 64\% (44\%) of the total without the 4428\,\AA\ DIB.
Together with the 4428\,\AA\ DIB, they contain most of the carbon that is included in the DIBs, probably in the form of cations (e.g.\ Ruiterkamp et al.\ 2005, Salama 2008). This may be  one or two orders  of magnitude more than in all the narrow DIBs  if their average f-value were significantly lower. 
For instance, assuming a low value for $\bar{f}$\,=\,0.06 (0.001 per C atom, see Section 6) with EW$_{\rm 1tot}$\,=\,0.9\,x\,10$^4$ would yield a significant  value for the interstellar carbon fraction X$_1$\,=\,1.2\,x\,10$^{-2}$,  
that is,\ $\sim$10\% of that in PAHs in HD\,183143.

The great majority of PAHs (cations) cannot contribute to the individually observed DIBs, but they must contribute to the part of the visible  interstellar extinction attributed to PAHs (e.g.\ Li \& Draine 2001) in the form of a continuum background of unresolved undetectable very weak DIBs, as explained in the following section.

\section{Background of unresolved very weak PAH DIBs}

Because  they are a significant component of interstellar dust, PAHs contribute to a few percent of the visible  interstellar extinction, as estimated by Weingartner \& Draine (2001), for example. In this model, we see from Figure 16 of Li \& Draine (2001) that the PAHs might contribute roughly to $\sim$6\%  of the standard total extinction at 6000\,\AA\  in the diffuse ISM of the Milky Way. 
From the uncertain values of the absorption cross section per C atom of ionised
PAHs of Figure 2 of Li \& Draine (2001), we  may infer that the average f-value per C-atom in the 5500-7000\,\AA\  range could be $\sim$0.001-0.002, which would yield an average f-value equal to 0.1-0.2 for a typical PAH with N$_{\rm C}$\,=\,100. This is compatible with the value of 0.1 proposed by Salama (2001) for the average f-value of PAH cations that must dominate the PAH extinction in the visible  range. 

Using the corresponding value $\bar{f}$\,=\,0.06 for N$_{\rm C}$\,=\,60 and X$_{\rm PAH}$\,=\,0.1 in Eq.\ (2), we find the formal total equivalent width E$_{\rm W1}$\,$\approx$\,1.3x10$^5$\,m\AA /mag for all PAHs in the visible  range in HD\,183143. After subtracting the PAH contribution to DIBs, $\sim$1-2\,x\,10$^4$\,m\AA /mag, we see that the total equivalent width of the PAH unresolved background, $\sim$10$^5$\,m\AA /mag, might be about five times that of all DIBs and ten times that of the very broad DIBs (not including the 4428\,\AA\ DIB) in Table 2. 
This suggestion is somewhat similar to the so-called grass concept, which is related to the number of unresolved DIBs per equivalent width, and was developed by Tielens (2014, Fig.\ 1; see also Gredel et al.\ 2011).

The number of unresolved very weak DIBs that contribute to this background probably is extremely high. First, considering a total spectral range of about 3000\,\AA , from $\sim$4500\,\AA\ to $\sim$7500\,\AA\ and a width $\sim$0.5-1\,\AA , there cannot be fewer than about 5000 bands that give an unresolved continuum. However, their equivalent width is expected to be smaller than the detection limit of actual DIBs, $\la$ 5\,m\AA /mag. This means that the total equivalent width of the PAH unresolved background, $\sim$10$^5$\,m\AA /mag, must be shared between more than 20,000 unresolved very weak DIBs, and probably by many more. The rapid multiplication of PAH isomers with their number of carbon atoms N$_{\rm C}$, plus hetero atoms, might well explain such large numbers. However, the increase in vibration modes must also play a role. For N$_{\rm C}$\,$\ga$\,200-300, the absorption spectrum of each PAH cation is itself expected to evolve toward some kind of continuum because a multitude of vibronic lines contribute to it (Salama et al.\ 1996).

\section{Conclusion}

The correlation between the wavelength of optical DIBs and the apparent UV resilience of their carriers is remarkable, although it does not apply to infrared DIBs, which seem to be mostly $\sigma$-type. As expected, we find that this correlation is based on the narrow DIBs, which form the most numerous DIB class, while it does not apply to the majority of the broader DIBs, although they contribute most of the total DIB absorption. The most prominent narrow DIBs belong to one of the two well-studied $\zeta$ or C$_2$ DIB families that are known to be sensitive to UV quenching and to present similar characteristic molecular band profiles. It is tempting to surmise that all narrow DIBs display similar profiles, but this requires confirmation by further sensitive DIB observations. It therefore seems likely that these DIBs are carried by molecular compounds that are similar to those that explain the observed $\lambda$/UV correlation. These carriers might be elongated systems of $\pi$ correlated electrons, such as carbon chains or catacondensed PAHs. These compounds may display series of strong optical bands that tightly depend on their length, as has been found  for chains and polyacenes. Their photodissociation also strongly depends on their length, which could explain the $\lambda$/UV correlation as proposed for polyacenes by OBT19.

The total amount of carbon contained in all the carriers of these narrow DIBs remains a very small fraction of the interstellar carbon if their oscillator strengths are $\ga$\,1. The amount of carbon  locked in the carriers of the broader DIBs could be much higher, especially if their oscillator strengths are significantly weaker.

The proposed identification of the carriers of significant interstellar absorption through the whole visible  range with simple conjugated systems of $\pi$ electrons that are related to the basis of most of our dyes 
might appear nice and natural. However, this conjecture remains fragile as long as not a single such DIB has been identified and the presence of such compounds in the diffuse ISM has not been confirmed. Among the variety of possible molecules, carbon chains and catacondensed PAHs with a single-hexagon chain such as polyacenes and derivatives, or polyrylenes seem to be favoured for their simplicity and symmetry and the presence of related species in the ISM.
Smaller chains are well known in dense molecular clouds and the rich PAH family is one of the main interstellar carbon reservoirs in galaxies. Polyacenes and related PAHs seem to better meet the observed $\lambda$/UV correlation than chains. However,  other origins for this correlation  are possible.

The PAH cationic bands might explain part of the  correlation between the DIB wavelength and the UV strength at large $\lambda$. It might seem attractive to link the multiplication of these weak UV-favoured DIBs with the contribution of PAHs to the visible  interstellar absorption if both are due to PAH cations with weak oscillator strengths. However, strong DIBs are much better explained by singular carriers with strong optical bands. Elongated PAHs could be appropriate candidates because of  their very strong optical bands. Some specific chemical processes are then required to compensate for their lower stability compared to the bulk of other PAHs. It is not excluded that the complex chemistry of grain surface and PAHs could offer such a chemical path similar to the easiness of the synthesis of graphene nanoribbons on surfaces. The complexity of such processes prevents any  definitive conclusion, and further studies would be worthwhile. 

Laboratory spectroscopy of long polyacenes (or polyrylenes) in gas phase or helium droplets appears as the best way to try to confirm them as DIB carriers.  Despite the problem of their instability, such studies do not appear impossible for already synthesised species such as decacene or undecacene, which have been proposed as possible carriers of C$_2$ DIBs (OBT19).

\bigskip

\begin{acknowledgements}
We thank the referee for his/her very helpful comments
and suggestions. We are indebted to the late Sydney Leach and Rosine Lallement  for many discussions and important  suggestions, and to Pierre Cox for his careful reading of the manuscript and his suggestions for improving it. We would like to thank Olivier Bern\'e, Christine Joblin, Benoit Soep, 
Elisabetta Micelottta and Christina T\"{o}nshoff  for discussions and comments on various aspects.  
We made a high use of "TOPCAT \& STIL: Starlink Table/VOTable Processing Software" (Taylor 2005) and thank its developers.
This work was supported in part by the German Research Foundation (DFG). 

\end{acknowledgements}

\bigskip

{\bf References}

Ajayan, P.~M., \& Yakobson, B.~I.\ 2006, Nature, 441


Allain, T., Leach, S., \& Sedlmayr, E.\ 1996a, \aap, 305, 602 

Allain, T., Leach, S., \& Sedlmayr, E.\ 1996b, \aap, 305, 616 

Battaglia, S., Faginas-Lago, N., Andrae, D., Evangelisti, S., \& Leininger, T.\ 2017, J.\ Phys.\ Chem.\ A, 121, 3746





Boschman, L., Cazaux, S., Spaans, M., Hoekstra, R., \& Schlath{\"o}lter, T.\ 2015, \aap, 579, A72

Cai, J., et al.\ 2010. Nature 466, 470

Cami, J., Sonnentrucker, P., Ehrenfreund, P., \& Foing, B.~H.\ 1997, \aap, 326, 822 


Cami, J.\ 2014,  The Diffuse Interstellar Bands, IAU Symposium, 297, 370 

Cami, J., \& Cox, N.~L.~J.\ 2014,  The Diffuse Interstellar Bands, IAU Symposium, 297

Cami, J., Cox, N.~L., Farhang, A., Smoker, J., Elyajouri, M., Lallement, R., et al.\ 2018, The Messenger, 171, 31 

Campbell, E.~K., Holz, M., Gerlich, D., \& Maier, J.~P.\ 2015, \nat, 523, 322 

Campbell, E.~K., Holz, M., \& Maier, J.~P.\ 2016, \apjl, 826, L4 

Campbell, E.~K., \& Maier, J.~P., 2017, J.\ Chem.\ Phys., 146, 160901

Castellanos, P., Candian, A., Andrews, H., \& Tielens, A.~G.~G.~M.\ 2018a, \aap, 616, A166 

Castellanos, P., Candian, A., Zhen, J., Linnartz, H., \& Tielens, A.~G.~G.~M.\ 2018b, \aap, 616, A167  

Chabot, M., Tuna, T., B{\'e}roff, K., et al.\ 2010, \aap, 524, A39 



Chen,T.,  Zhen, J., Wanga, Y.,  Linnartz, H., \&  Tielens, A.~G.~G.~M.\ 2018, Chem.\ Phys.\ Lett., 
692, 298

Cox, N.~L.~J., Cami, J., Kaper, L., et al.\ 2014, \aap, 569, A117

Dewar, M.~J.~S., \& Longuet-Higgins, H.~C.\ 1954 Proc.\ Phys.\ Soc., A 67 795

Douglas, A.~E.\ 1977, Nature, 269, 130 

Duley, W.~W.\ 2000, \apj, 528, 841 

Elyajouri, M., Monreal-Ibero, A., Remy, Q., et al.\ 2016, \apjs, 225, 19

Elyajouri, M., Cox, N.~L.~J., \& Lallement, R.\ 2017, \aap, 605, L10

Elyajouri, M., Lallement, R., Cox, N.~L.~J., et al.\ 2018, \aap, 616, A143 

Ensor, T., Cami, J., Bhatt, N.~H., \& Soddu, A.\ 2017, \apj, 836, 162 

Fan, H., Welty, D.~E., York, D.~G., et al.\ 2017, \apj, 850, 194 

Fan, H., Hobbs, L.~M., Dahlstrom, J.~A., et al.\ 2019, \apj, 878, 151

Fischer, G., \& Maier, J.~P.\  1997, Chemical Physics, 223, 149

Foing, B.~H., \& Ehrenfreund, P.\ 1994, \nat, 369, 296

Foing, B.~H., \& Ehrenfreund, P.\ 1997, \aap, 317, L59

Galazutdinov, G.~A., Lee, J.-J., Han, I., et al.\ 2017, \mnras, 467, 3099

Geballe, T.~R., Najarro, F., Figer, D.~F., Schlegelmilch, B.~W., \& de La Fuente, D.\ 2011, \nat, 479, 200 

Geballe, T.~R., Najarro, F., de la Fuente, D., et al.\ 2014, IAU Symposium, 303, 75 

Gredel, R., Carpentier, Y., Rouill{\'e}, G., et al.\ 2011, \aap, 530, A26

Halasinski, T.~M.,  Weisman, J.~L.,  Ruiterkamp, R., et al.\ 2003, J.\ Phys.\ Chem.\ A, 107, 3660

Hamano, S., Kobayashi, N., Kondo, S., et al.\ 2016, \apj, 821, 42

Hamano, S., Kobayashi, N., Kawakita, H., et al.\ 2018, Bulletin de la Societe Royale des Sciences de Liege, 87, 276

Herbig, G.~H.\ 1975, \apj, 196, 129 

Herbig, G.~H.\ 1995, \araa, 33, 19 

Hobbs, L.~M., York, D.~G., Snow, T.~P., et al.\ 2008, \apj, 680, 1256 

Hobbs, L.~M., York, D.~G., Thorburn, J.~A., et al.\ 2009, \apj, 705, 3

Huang, J., \& Oka, T.\ 2015, Molecular Physics, 113, 2159 


Hudgins, D.~M., Bauschlicher, C.~W., Jr., \& Allamandola, L.~J.\ 2005, \apj, 632, 316

Jensen, W.~B.\ 2013, in Strom and Wilson; 
https://pubs.acs.org/doi/pdf/10.1021/bk-2013-1122.ch004
{\it Pioneers of Quantum Chemistry}

Joblin, C., Maillard, J.~P., D'Hendecourt, L., et al.\ 1990, \nat, 346, 729

Johansson, K.~O., et al.\ 2018, Science 361, 997

Jones, A.~P.\ 2016, Royal Society Open Science, 3, 160223 

Ka{\'z}mierczak, M., Schmidt, M.~R., Bondar, A., \& Kre{\l}owski, J.\ 2010, \mnras, 402, 2548 

Kos, J., Zwitter, T., Grebel, E.~K., et al.\ 2013, \apj, 778, 86

Kosynkin, D.~V., A Higginbotham, A.~L., Sinitskii, A., et al.\ 2009, Nature, 458, 07872

Krelowski, J., \& Walker, G.~A.~H.\ 1987, \apj, 312, 860

Kre{\l}owski, J., Galazutdinov, G., Godunova, V., et al.\ 2019, \actaa, 69, 159 

Lan, T.-W., M\'enard, B., \& Zhu, G.\ 2015, MNRAS, 452, 3629


 Li, A., \& Draine, B.~T.\ 2001, \apj, 554, 778

Li, J-L., Kudin,K.~N., McAllister, M.~J., et al.\ 2006, Phys.\ Rev.\ Lett., 96, 176101



Malloci, G., Mulas, G., Cappellini, G., \& Joblin, C.\ 2007, Chemical Physics, 340, 43 

Malloci, G., Cappellini, G., Mulas, G., \& Mattoni, A.\ 2011, Chemical Physics, 384, 19


Mondal, R., T\"{o}nshoff, C.,  Khon, D., et al.\ 2009, J.\ Am.\ Chem.\ Soc.,  131, 14281

Nakada, K., Fujita, M., Dresselhaus, G., \& Dresselhaus, M.~S.\ 1996, Phys.\ Rev.\ B, 54, 17954 

Oka, T., Welty, D.~E., Johnson, S., et al.\ 2013, \apj, 773, 42; 2014, \apj, 793, 68


Omont, A.\ 2016, \aap, 590, A52 

Omont, A., Bettinger, H.~F., \&  T\"{o}nshoff, C.\ 2019,  \aap, 625, 441 {\bf OBT19}



Pizzarello, S.\ \& Shock, E.\ 2010, Cold Spring Harb.\ Perspect.\ Biol., 2, a002105


Platt, J.~R. 1956, J.\ Chem.\ Phys., 25, 80




Puspitarini, L., Lallement, R., Babusiaux, C., et al.\ 2015, \aap, 573, A35

Rice, C.~A., \& Maier, J.~P.\ 2013, J.\ Phys.\ Chem. A, 117, 5559

Ruffieux, P., Wang, S., Yang, B., Sanchez, C., et al.\ 2016, Nature, 531, 489

Ruiterkamp, R., Halasinski, T., Salama, F., et al.\ 2002, \aap, 390, 1153 

Ruiterkamp, R., Cox, N.~L.~J., Spaans, M., et al.\ 2005, \aap, 432, 515

Salama, F., Bakes, E.~L.~O., Allamandola, L.~J., \& Tielens, A.~G.~G.~M.\ 1996, \apj, 458, 621


Salama, F.\ 2001, Journal of Molecular Structure, 563, 19

 Salama, F.\ 2008, Organic Matter in Space, 251, 357


Salama, F., \& Ehrenfreund, P.\ 2014,  The Diffuse Interstellar Bands, IAU Symposium, 297, 364 

Sanchez-Sanchez, C., et al.\ 2019 J.\ Am.\ Chem.\ Soc., 141, 3550

Sarre, P.~J.\ 2006, Journal of Molecular Spectroscopy, 238, 1 

Sarre, P.~J.\ 2014, The Diffuse Interstellar Bands, IAU Symposium, 297, 34 




Shen, B., Tatchen, J., Sanchez-Garcia, E., \& Bettinger, H.~F.\  2018, Angew. Chem. Int., 57, 10506

Simon, A., Champeaux, J.~P., Rapacioli, M., et al.\ 2018, Theoretical Chemistry accounts, 137:106





Taylor, M.~B.,  2005, Astronomical Data Analysis Software and Systems XIV, eds. P
Shopbell et al., ASP Conf. Ser. 347, 29

Thaddeus, P.\ 1995, in Tielens, A.~G.~G.~M., \& Snow, T.~P.\ 1995, The Diffuse Interstellar Bands, {\it Astrophysics and Space Science Library}, 202, Kluwer

Thorburn, J.~A., Hobbs, L.~M., McCall, B.~J., et al.\ 2003, \apj, 584, 339

Tielens, A.~G.~G.~M., \& Snow, T.~P., 1995. The Diffuse Interstellar Bands, {\it Astrophysics and Space Science Library}, 202, Kluwer

Tielens, A.~G.~G.~M.\ 2005, The Physics and Chemistry of the Interstellar Medium (Cambridge University Press, Cambridge, UK)

Tielens, A.~G.~G.~M.\ 2008, \araa, 46, 289 

Tielens, A.~G.~G.~M.\ 2013, Reviews of Modern Physics, 85, 1021

Tielens, A.~G.~G.~M.\ 2014,  The Diffuse Interstellar Bands, IAU Symposium, 297, 399

T{\"o}nshoff, C., \& Bettinger 2010, Angew.\ Chem., Int.\ Ed., 49, 4125

Vos, D.~A.~I., Cox, N.~L.~J., Kaper, L., Spaans, M., \& Ehrenfreund, P.\ 2011, \aap, 533, A129 

Weingartner, J.~C., \& Draine, B.~T. 2001, \apj, 548, 296

Weisman, J.~L., Lee, T.~J., Salama, F., \& Head-Gordon, M.\ 2003, \apj, 587, 256 

Welty, D.~E.\ 2014, The Diffuse Interstellar Bands, 153

Welty, D.~E., Ritchey, A.~M., Dahlstrom, J.~A., et al.\ 2014, \apj, 792, 106

Wu, C., Lee, P., \& Chai, J.\ 2016, Scientific Reports, 6, 37249


Zack, L.~N., \& Maier, J.~P.\ 2014a,  The Diffuse Interstellar Bands,  IAU Symposium, 297, 237

Zack, L.~N., \& Maier, J.~P.\ 2014b, Chem.\ Soc.\ Rev., 2014, 43, 4602

Zdetsis, A.~D., \& Economou E.~N.\ 2017, Carbon, 116, 422

Zeng, W., Qi, Q., \& Wu, J.\ 2018, Eur.\ J.\ Org.\ Chem., 2018, 7


Zuzak, R., Dorel, R.,  Kolmer, M., Szymonski, M., Godlewski, S., \& Echavarren, A. M.\ 2018, Angew. Chem. Int. Ed., 57, 10500

\appendix

\section{Influence of the definition of "narrow" DIBs}

This appendix describes the small differences in the properties of the $\lambda$/R$_{21}$ correlation with an alternative choice of 1.0\,\AA\ for the limit between narrow and broad DIBs. This is expressed in Table A.1 and Figs.\ A.1 to A.3, which correspond to Table 2 and Figs.\ 3 to 5 with this new limit. A linear fit gives
\begin{equation}
 {\rm logR_{21}\,=\, -3.77\,\times\,10^{-4}\times\lambda(A)\,+\,2.23 \pm 0.28}
\end{equation}
 with a Pearson-correlation coefficient  P\,=\,-0.62 for the narrow DIBs (Fig. A.1). These fit and Pearson coefficients are almost identical to those of Fig.\ 3 and Eq.\ (1) for a limit of 1.1\,\AA .

\begin{table}[htbp]
      \caption[]{Properties of narrow and broad DIBs detected on the lines of sight of HD\,183143 (1) and HD\,204827 (2), derived when a limit of 1.0\,\AA\  between the two classes is adopted.}
         \label{tab:lines}
            \begin{tabular}{ l c c  c c    }
            \hline 
           \noalign{\smallskip}
  &   1 and 2$^a$  &    1  not 2$^b$ & 2  not 1$^c$ & {\tiny Total} \\
      \noalign{\smallskip}
  Narrow {\tiny($<$1.0\AA )} & & & &  \\       
{\small Number}   &  207  &  67 &  82  & 356  \\
EW$_{\rm 1av}$ & 17.1 &   9.6  &   x  &  x     \\  
EW$_{\rm 1tot}$ & 3540 &   640  &  x  & 4180 \\  
EW$_{\rm 2av}$ & 11.2  & x &   6.1    &  x    \\  
EW$_{\rm 2tot}$ & 2320 & x &  500     & 2820 \\  
          \hline
  & & & &  \\
Broad$^d$ {\tiny($>$1.0\AA )} & & & &  \\       
{\small Number}  & 60   &  83 &  9   &  152 \\
EW$_{\rm 1av}$ & 120 &   32.9  & x  &   x     \\  
EW$_{\rm 1tot}$ & 7200 &  2700  & x &9900 \\  
EW$_{\rm 2av}$ & 43     & x  &  21.7     &   x     \\  
EW$_{\rm 2tot}$ &2580  & x & 200    & 2780  \\ 
          \hline
  & & & &  \\
4428\,\AA & & & & \\
EW$_1$ & 5700 & x &x & 5700 \\
EW$_2$ & 1220 & x &x & 1220  \\
      \noalign{\smallskip}
            \noalign{\smallskip}
            \hline
           \end{tabular}
See notes of Table 2.
\end{table}

\begin{figure}[htbp]
         \begin{center}
\includegraphics[scale=0.63, angle=0]{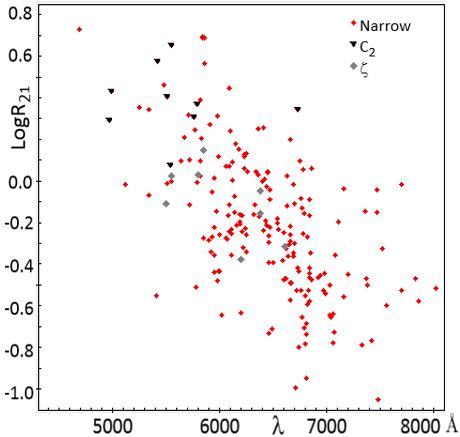}
\caption{Wavelength dependence of the ratio R$_{21}$ of the equivalent widths of narrow DIBs in the lines of sight of  HD\,204827 and HD\,183143 (Hobbs et al.\ 2008, 2009)  confirmed in Fan et al.\ (2019), derived when a limit of 1.0\,\AA\ between narrow and broad DIBs is adopted (Table A.1). These DIBs are shown by red dots, except for the DIBs identified in Table 1 as $\zeta$ (Ensor et al.\ 2017; grey diamonds) or C$_2$ (Thorburn et al.\ 2003; black triangles).} 
     \end{center}
 \end{figure}

\begin{figure}[htbp]
         \begin{center}
\includegraphics[scale=0.57, angle=0]{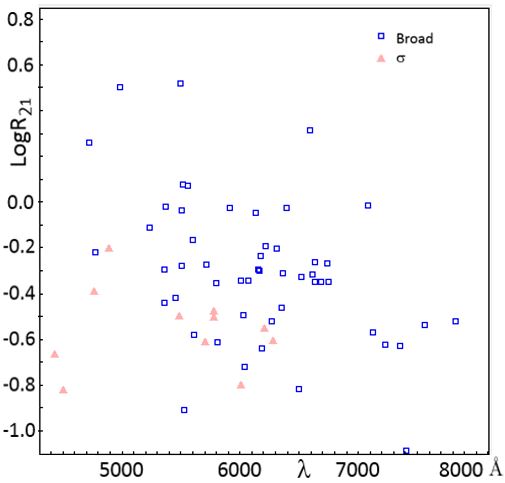}
 \caption{Same as Fig.\ A.1 for broad DIBs (FWHM\,$>$\,1.0\,\AA , Table A.1). These DIBs are shown by blue open squares, except for the DIBs identified in Table 1 as $\sigma$ (Lan et al.\ 2015) which appear as orange full triangles.}
     \end{center}
 \end{figure}

\begin{figure}[htbp]
         \begin{center}
\includegraphics[scale=0.58, angle=0]{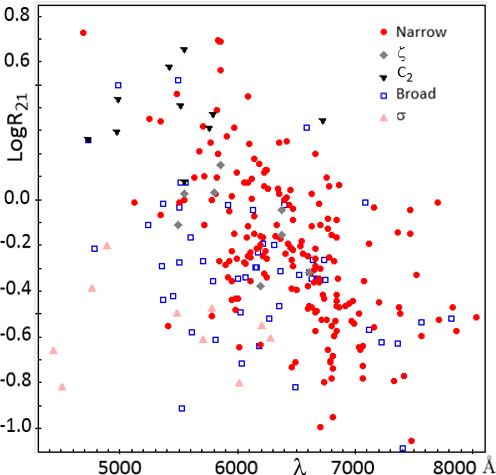}
 \caption{Same as Fig.\ A.1 for all DIBs (same as Fig.\ 4 of OBT19 and Fig.\ 5) with the same symbols as in Figs.\ A.1 \& A.2.}
     \end{center}
 \end{figure}

\section{Expected dependence of R$_{21}$ on $\lambda$ for polyacenes}

As shown in Section 3.1, Eq(1), the wavelength dependence of the DIB intensity ratio, R$_{21}$\,=\,EW(204827)/EW(183143), in the sightlines HD\,183143 and HD\,204827 (Fig.\ 3) may be fitted by the exponential relation 

\begin{equation}
 {\rm logR_{21}\,=\, -3.7\,\times\,10^{-4}\times\lambda(A)\,+\,2.2\,\pm\,0.33}
.\end{equation}

In this appendix we explore  the possible origin of this  dependence in more detail  if the DIB carriers are polyacenes or other elongated molecules with a similar series of strong optical bands. As stated in OBT19 and Section 1, this dependence may be qualitatively expected from  the strong dependence of the H photolysis of polyacenes on their length, while there is a linear progression of the wavelength of the strong optical band of polyacenes with this length, as shown by Figure 2 of OBT19, which may be extrapolated as 

\begin{equation}
 {\rm \lambda (A) \approx  100\,\times  N_{C} + 200 = 400\,\times  N + 400}
,\end{equation} 

where the approximate argon-matrix shift, $\sim$200\,\AA , used in OBT19 is included and N$_{\rm C}$ = 4\,N + 2 is the number of carbon atoms of the polyacene with N rings.

As recalled in OBT19, H photolysis is thought to be the main destruction process of polyacenes for the relevant range of N $\sim$ 10-18. Their abundances and the associated DIBs are therefore expected to be determined by the balance between H photolysis and H accretion onto their dehydrogenated counterparts, as in the standard theory of PAH photochemistry (e.g.\ Tielens 2005, 2013). The most important parameter for determining the polyacene abundances versus N or N$_{\rm C}$ and  the UV-radiation properties is the photodissociation probability, following a UV-photon absorption. This probability is proportional to exp(-E$_0$/kT$_{\rm v}$) (see e.g.\ Eq.\ (45) of Tielens 2013), where E$_0$ is related to the H binding energy and T$_{\rm v}$ is the internal vibrational temperature. T$_{\rm v}$ is related to N$_{\rm C}$ through the approximate relation (Tielens 2005, Eq.\ 3)
\begin{equation}
 {\rm T_{\rm v} = 2000 \times [E_{UV}(eV)/N_C]^{0.4}
}
.\end{equation}
For a typical value E$_{\rm UV}$\,=\,10\,eV of the energy of the absorbed UV photon in unshielded diffuse clouds, this yields T$_{\rm v}$ = 1130\,K and 940\,K for N\,=\,10 and 16 (N$_{\rm C}$\,=\,42 and 66), respectively. This corresponds  to $\lambda$\,=\,4400\,\AA\ and 6800\,\AA\ from Eq.\ (B.2), respectively.

When we assume E$_0$\,=\,3.52\,eV ($\sim$41\,000\,K) as in Andrews et al.\ (2016), for example, the same values of N$_{\rm C}$ yield 
E$_0$/kT$_{\rm v}$\,$\sim$\,36 and 44, and exp(E$_0$/kT$_{\rm v}$)\,$\sim$\,6x10$^{15}$ and 10$^{19}$, respectively, which is\ a variation by a factor $\sim$1500 through the optical range $\sim$4400-6800\,\AA . This is two orders of magnitude more than the variation by a factor $\sim$8  of R$_{21}$ given by Eq(1). While it is probable that this exponential dependence of the PAH photodissociation rate is the origin of the observed $\lambda$/UV correlation, the latter can therefore not be proportional to this exponential factor. 

Like other PAHs, the dehydrogenation of polyacenes varies extremely sharply  around a critical value of N$_{\rm C}$ corresponding to a critical value  of N,  N$_{\rm cr}$, so that the number of fully hydrogenated polyacenes abruptly decreases to zero below N$_{\rm cr}$ (OBT19). As seen from Eq.\ (B.3),  the critical parameter for photodissociation is indeed N$_{\rm C}$/E$_{\rm UV}$, so that N$_{\rm cr}$ and the corresponding $\lambda_{\rm cr}$ must be roughly  inversely proportional  to E$_{\rm UV}$. 

While the energies of UV photons are always distributed, the PAH photodissociation is dominated by the hardest UV photons around an effective E$_{\rm UV}^{eff}$. In unshielded diffuse clouds, such as on the sightline of HD\,183143, E$_{\rm UV}^{eff}$ is expected to be $\sim$10-13\,eV. Its value is substantially lower in UV-shielded regions, such as are found on the sightline of HD\,204827, possibly by a factor up to $\sim$1.5. N$_{\rm cr}$ and $\lambda_{\rm cr}$ are  expected to be reduced by a similar factor. This might explain why the DIBs with short wavelength are stronger and more numerous in HD\,204827 than in HD\,183143. This explanation seems compatible with the observed variation of R$_{21}$ with $\lambda$ (Fig.\ 3). However, a precise modelling of the variation of R$_{21}$ with $\lambda$ aiming at quantitatively accounting the observed correlation of Eq.\ (1) currently appears to be beyond reach because of the various sources of complexity, including 1) the uncertain nature of the possible elongated DIB carriers and their optical spectrum; even  the spectrum of neutral polyacenes is still uncertain for N\,$>$\,11. 2) The complexity of the photochemistry of PAHs. 3) The complexity of the structure of the interstellar medium along any line of sight, inducing much uncertainty about the properties of the interstellar UV radiation field and its radiative transfer. In particular, it is certain that the sightline of HD\,204827 is not entirely strongly shielded but has unshielded parts, at least at the boundary of denser clumps. It is also possible that HD\,183143 has a few clumps that are significantly denser than the average. 

 This mixed character of the sightlines might  explain a good part of the correlation. We assume, for instance, that i) the sightline of HD\,204827 includes 60\% well-shielded regions that allow the inattenuate abundance of fully hydrogenated acenes down to N\,$\sim$\,11 ($\lambda$\,$\sim$\,4800\,\AA ) and 40\% unshielded regions similar to most of the sightline of HD\,183143. We also assume that ii) the sightline of HD\,183143 includes 10\% well-shielded regions and 90\% unshielded regions where polyacenes are fully dehydrogenated for N\,$\la$\,12  ($\lambda\,\la$\,5500\,\AA ) and fully hydrogenated for N\,$\ga$\,15 ($\lambda\,\ga$\,6500\,\AA ). Based on this, we expect for the DIB ratio R$_{21}$\,=\,EW(204827)/EW(183143)
\begin{equation}
 {\rm R_{21}(5000\,A)\approx\,(60/10)\times K\times \chi_{2T} (N11)/\chi_{1T} (N11)
}
\end{equation}
\begin{equation}
 {\rm R_{21}(7000\,A) \approx   K \times \chi_{2T} (N16)/\chi_{1T} (N16)
}
,\end{equation}

where $\chi_{1T}$(N) ($\chi_{2T}$(N)) is the total abundance of various hydrogenation states of N-acene in the sightline of HD\,183143 (HD\,204827) and K is a constant. If the ratio $\chi_{2T}$(N)/$\chi_{1T}$(N) does not vary strongly with N in this N range, this  yields  
\begin{equation}
 {\rm R_{21}(\sim 5000\,A) / R_{21}(\sim 7000\,A) \approx 6}
.\end{equation}
This result is very close to the value, 5.5, given by Eq.\ (1) on the same wavelength interval for the fit of the observed correlation. This agreement is not very significant considering the uncertainties on the structure of the lines of sight. It nevertheless appears to confirm that the optical properties of large polyacenes or similar molecules might account for most of the observed R$_{21}$/$\lambda$ correlation. Moreover, the large number of cationic bands expected at high $\lambda$ may increase the steepness of the R$_{21}$/$\lambda$ relation.

\end{document}